\documentclass[%
 reprint,
nofootinbib,
amsmath,amssymb,
aps,
prd,
superscriptaddress]{revtex4-2}
\allowdisplaybreaks[3]
\usepackage{graphicx,subfigure}
\usepackage{dcolumn,tabularx,calc,multirow,booktabs,slashed,soul,color}
\usepackage{bm}
\usepackage{srcltx}
\usepackage[colorlinks=true,citecolor=blue,urlcolor=cyan,filecolor=magenta,linkcolor=red,frenchlinks=true,bookmarks]{hyperref}

\usepackage{ulem,soul}

\usepackage{xcolor}

\begin{document}

\preprint{APS/123-QED}

\title{Generalizing the Soffer Bound: Positivity Constraints on Parton Distributions of Spin-3/2 Particles}

\author{Dongyan Fu}~\email{fudongyan@impcas.ac.cn}

\affiliation{Southern Center for Nuclear-Science Theory (SCNT), Institute of Modern Physics, Chinese Academy of Sciences, Huizhou 516000, China}

\author{Yubing Dong}~\email{dongyb@ihep.ac.cn}

\affiliation{Institute of High Energy Physics, Chinese Academy of Sciences, Beijing 100049, China}
\affiliation{School of Physical Sciences, University of Chinese Academy of Sciences, Beijing 101408, China}

\author{S. Kumano}~\email{kumanos@impcas.ac.cn}

\affiliation{Southern Center for Nuclear-Science Theory (SCNT), Institute of Modern Physics, Chinese Academy of Sciences, Huizhou 516000, China}
\affiliation{Quark Matter Research Center, Institute of Modern Physics, Chinese Academy of Sciences,
Lanzhou, 730000, China}
\affiliation{KEK Theory Center, Institute of Particle and Nuclear Studies, KEK,
Oho 1-1, Tsukuba, 305-0801, Japan}

\author{Ju-Jun Xie}~\email{xiejujun@impcas.ac.cn}

\affiliation{Southern Center for Nuclear-Science Theory (SCNT), Institute of Modern Physics, Chinese Academy of Sciences, Huizhou 516000, China}
\affiliation{Heavy Ion Science and Technology Key Laboratory, Institute of Modern Physics, Chinese Academy of Sciences, Lanzhou 730000, China}
\affiliation{School of Nuclear Sciences and Technology, University of Chinese Academy of Sciences, Beijing 101408, China}

\begin{abstract}
We derive the complete set of positivity bounds for the leading-twist parton distribution functions (PDFs) of a spin-3/2 hadron for the first time. 
This work generalizes the Soffer bound, a fundamental constraint for spin-1/2 nucleons, to quark and gluon distribution functions in higher-spin systems. 
Expressing the antiparton-hadron scattering amplitudes in terms of the PDFs and the spin density matrix, we establish the connections between the PDFs and the scattering amplitudes in the tensor product space of the parton and hadron spins.
Moreover, we obtain the definitions of the PDFs in terms of the helicity amplitudes.
Positive definiteness of the scattering amplitude matrix yields a set of inequalities that define the physically allowed parameter space for the helicity amplitudes and a set of constraints for the PDFs.
The Cauchy-Schwartz inequality determines the constrains between the PDFs and generalized parton distributions.

\end{abstract}

\maketitle

\paragraph*{Introduction.}
The internal structure of hadrons, described by the parton distribution functions (PDFs), is fundamental to understand strong interaction dynamics in Quantum Chromodynamics (QCD).
For spin-1/2 hadrons, the theoretical framework is well-established, with three leading-twist quark distributions—the unpolarized distribution \( f_1(x) \), the longitudinal distribution \( g_1(x) \), and the transversity distribution \( h_1(x) \) and there have been many studies~\cite{Collins:1981uw,Soffer:1994ww,Vogelsang:1997ak,AsymmetryAnalysis:1999gsr,Pobylitsa:2001nt,Kirch:2005in,DAlesio:2020vtw,Cocuzza:2023vqs,Jahan:2025sgx}.
A critical theory constraint on these functions is the Soffer bound, \( |h_1(x)| \leq \frac{1}{2} [f_1(x) + g_1(x)] \)~\cite{Soffer:1994ww,Pire:1998nw}, which is derived from the positivity of probability densities and provides essential constraints for global QCD analyses of the polarized deep-inelastic scattering and provides a fundamental test method for the self-consistency of the parton model~\cite{Maji:2018fap,Constantinou:2020hdm,Cheng:2023kmt,Lorce:2025aqp}.

The extension of this formalism to hadrons with higher spin presents a natural and important frontier in hadron structure physics. As the particle spin increases, there are additional PDFs, like the higher order unpolarized-parton PDF \( f_{1LL} (x)\) for quark in the spin-1 particle~\cite{Bacchetta:2001rb,Cotogno:2017mwy} and longitudinal-parton PDF \(g_{1LLL}(x)\) and transversity-parton PDF \(h_{1LLT}(x)\) in the spin-3/2 particle. The gluon PDFs have the similar property. For these systems, the basis of the PDFs is expanded by additional functions related to vector and tensor polarizations. The generalization of positivity constraints to spin-1 particles was systematically derived and established a more complex set of inequalities that must be satisfied by their PDFs~\cite{Bacchetta:2001rb,Keller:2022abm,Kumano:2024fpr,Kumano:2025rai}. 

For spin-3/2 hadrons like the \(\Delta(1232)\) resonance, the richer polarization information leads to the new PDFs.
Formally defining these distributions requires careful treatment within the Rarita-Schwinger formalism~\cite{Rarita:1941mf}. Several theoretical studies have explored aspects of the spin-3/2 hadron structure and our early work established formal definitions for some structure related functions, like generalized parton distributions (GPDs)~\cite{Fu:2022bpf,Fu:2024kfx}.
Despite this progress, a systematic derivation of the complete set of positivity bounds for the spin-3/2 PDFs, a direct and necessary generalization of the spin-1/2 Soffer bound and the spin-1 constraints, has not been presented. Such bounds are indispensable. They ensure the physical viability of model predictions, provide critical constraints for future global QCD fits, and are prerequisites for the reliable extraction of these functions from experimental data or lattice QCD calculations~\cite{Chai:2020nxw}.

In this work, from the correlation function that defines the PDFs, we give the antiparton-hadron scattering amplitude in terms of the PDFs of the spin-3/2 particle. For the quark and gluon, we respectively derive the relations between the PDFs and helicity amplitudes, and obtain the constraints on the PDFs according to the positivity of the scattering amplitude.
In the non-forward limit, we also derive the bounds between the PDFs and GPDs.

\paragraph*{PDFs of the spin-3/2 particle.}
The PDFs of the quark and gluon can be respectively defined from the light-cone correlation functions,
\begin{eqnarray}
\Phi_{ij,ab}^q(x) &=& \int \langle P, S, T, R | \bar{\psi}_{i,b}(0) \psi_{j,a}(z^-)| P, S, T, R \rangle, \label{correlation-function-quark} \\
		\Phi_{ij}^{g, \mu \nu}(x) &=& \frac{1}{P^+} \int \langle P, S, T, R | G_{i}^{+ \mu}(0) G_{j}^{\nu +}(z^-) |P, S, T, R \rangle, \label{correlation-function-gluon}
\end{eqnarray}
where $\int$ represents $\int \frac{\text{d} z^-}{2 \pi} e^{i x P^+ z^-}$ and also in the following content, $x$ is the light cone momentum fraction of the parton, $x = k^+/P^+$, in which $k$ and $P$ respectively are the momenta of the parton and the target hadron. 
The convention of the four-vector $v$ in the light cone coordinate is $v = (v^+, v^-, \bm{v}_T)$, with $v^{\pm} = \frac{1}{\sqrt{2}}(v^0 \pm v^3)$ and $\bm{v}_T = (v^1, v^2)$.
In the following, the vector carrying the subscript $T$ implies that the corresponding index only takes $1, 2$ or $x, y$ components.
The $\psi$ and $G^{\rho \delta}$ are the quark field and gluon field strength tensor.
The indices $i$ and $j$ represent the helicities of the absorbed and emitted partons, with $i,j = \pm \frac{1}{2}$ for quark and $i,j = \pm 1$ for gluon due to the different spin.
The $a$ and $b$ are the 4-component indices of the spinors, and all the traces in the following act on $a$ and $b$. In Eqs.~\eqref{correlation-function-quark} and \eqref{correlation-function-gluon}, the $S$, $T$, and $R$ carry the polarization information of the spin-3/2 hadron.
In this work, the light-cone gauge is considered, so that the Wilson line does not appear in the correlation functions.

For the spin-3/2 particle, one can specify the polarization state by the spin vector $S$, the rank-2 spin tensor $T$ and the rank-3 spin tensor $R$ in the target rest frame~\cite{Zhao:2022lbw}.
By considering only the leading twist, the quark matrix $\Phi^q \gamma^+$ contains all the quark PDF information,
\begin{eqnarray}\label{correlator}
\Phi^q\gamma^+ &= & \left( f^q_1 + g^q_1 S_L \gamma_5 + h^q_1 \gamma_5 \slashed{S}_T + f^q_{1LL} S_{LL} \right. \nonumber \\
        & & \left. + g^q_{1LLL} S_{LLL}\gamma_5  + h^q_{1LLT} \gamma_5 \slashed{S}_{LLT} \right) \mathcal{P}_+,
\end{eqnarray}	
where $\mathcal{P}_+ \equiv \frac{1}{2} \gamma^- \gamma^+$ is the projection operator, which projects the fermion field to the ``good" component, $\psi_+\equiv \mathcal{P}_+ \psi$.
Meanwhile, the gluon correlation function is decomposed to the leading twist PDFs as,
\begin{equation}
    \begin{split}
        \Phi^{g, \mu \nu} =& \frac12 \left( f_1^g g_T^{\mu \nu} + g_1^g S_L i \epsilon_T^{\mu \nu} + h_{1TT}^g S_{TT}^{\mu \nu} \right. \\
        & \left. + f_{1LL}^g S_{LL} g_T^{\mu \nu} + g_{1LLL}^g S_{LLL} i \epsilon_T^{\mu \nu} \right),
    \end{split}
\end{equation}
where
\begin{eqnarray}
        g_T^{\mu \nu} & \equiv & g^{\mu \nu} - n^{\mu}\bar{n}^{\nu} - n^{\nu}\bar{n}^{\mu}, \\
        \epsilon_T^{\mu \nu}  &\equiv& \epsilon_{\alpha \beta \rho \sigma} g_T^{\alpha \mu} g_T^{\beta \nu} \bar{n}^{\rho} n^{\sigma}, \\
        \tau_{T,\mu \nu; \rho \sigma}^{} & \equiv & \frac{1}{2} g_{T \mu \rho} g_{T \nu \sigma} + \frac{1}{2} g_{T \mu \sigma} g_{T \nu \rho} - \frac{1}{2} g_{T \mu \nu} g_{T \rho \sigma},
\end{eqnarray}
with $n^\mu = 1/\sqrt{2}\,(1,0,0,-1)$ and $\bar{n}^{\mu} = 1/\sqrt{2}\,(1,0,0,1)$.

Moreover, the correlation function $\Phi(x)$ has the following properties under Hermite, parity and time-reversal transformations,
    \begin{eqnarray}
	\Phi^{q \dagger}(k,P,S,T,R|n) &= & \gamma^0 \Phi^q(k,P,S,T,R|n)\gamma^0,\\
	\Phi^q(k,P,S,T,R|n) \!\! &= & \!\! \gamma^0 \Phi^q(\bar{k},\bar{P},-\bar{S},\bar{T},-\bar{R}|\bar{n}) \gamma^0,\\
	\Phi^{q*}(k,P,S,T,R|n) &= & (-i \gamma_5 C) \Phi^q(\bar{k},\bar{P},\bar{S},\bar{T},\bar{R}|\bar{n}) \nonumber \\
 && \times (-i \gamma_5 C),
\end{eqnarray}
for quark~\cite{Kumano:2020ijt}, and
    \begin{eqnarray}
&& \!\!	\Phi^{g, \mu \nu *}(k,P,S,T,R|n) = \Phi^{g, \nu \mu}(k,P,S,T,R|n),\\
&&	\Phi^{g, \mu \nu}(k,P,S,T,R|n)  =  \Phi^{g, \mu \nu}(\bar{k},\bar{P},-\bar{S},\bar{T},-\bar{R}|\bar{n}),\\
&& \Phi^{g, \mu \nu*}(k,P,S,T,R|n) =  \Phi^{g, \mu \nu}(\bar{k},\bar{P},\bar{S},\bar{T},\bar{R}|\bar{n}),
\end{eqnarray}
for gluon~\cite{Mulders:2000sh,Boer:2016xqr}, where the charge conjugation is $C=i \gamma^2 \gamma^0$, the polarization vector and tensor are transformed under the Lorentz covariant forms~\cite{Zhao:2022lbw}.
To explain the transformation $\bar{X}$, one can define the tensor $\bar{g}^\mu_{\,\,\, \alpha} = \operatorname{diag}(1,-1,-1,-1)$, which will add a minus sign to the 3-components of the 4-components. 
Then, the vector and tensor are defined by
\begin{equation}
		\bar{v}^\mu = \bar{g}^\mu_{\,\,\, \alpha} v^\alpha, \,\, \bar{T}^{\mu \nu} = \bar{g}^\mu_{\,\,\, \alpha} \bar{g}^\nu_{\,\,\, \beta} T^{\alpha \beta},\,\,
		\bar{R}^{\mu \nu \rho} = \bar{g}^\mu_{\,\,\, \alpha} \bar{g}^\nu_{\,\,\, \beta} \bar{g}^\rho_{\,\,\, \delta} R^{\alpha \beta \delta},
\end{equation}
where vector $v$ stands for $k$, $P$ and $S$.
One finds that the correlation function $\Phi(x)$ satisfies the Hermite, parity and time-reversal transformations.

\begin{table}[htbp]
	\renewcommand\arraystretch{1.3}
	\caption{\small{Leading twist projection operator $\Gamma_{i j}$ to different parton helicity components.}}\vspace{0.5em}
	\centering
 \scalebox{0.9}{\begin{tabular}{ c c c c c c c}
		 \toprule
		 \toprule
		   $ij$ & $++$ & $- -$ & $- +$ & $+ -$ \\
		\midrule
		$\Gamma^q_{i j}$ & $\frac{1}{4} (1+\gamma_5)$	& $\frac{1}{4} (1-\gamma_5)$	& $\frac{1}{4} (\gamma^1-i \gamma^2)$	& $-\frac{1}{4} (\gamma^1 + i \gamma^2)$ \\
        \midrule
		$\Gamma^g_{i j, \mu \nu}$ & $ \frac12 (g_{\mu \nu} - i \epsilon_{T \mu \nu})$	& $\frac12 (g_{\mu \nu} + i \epsilon_{T \mu \nu})$	& $\frac12 \tau_{T, \bar{c} \bar{c}, \mu \nu}$	& $\frac12 \tau_{T, c c, \mu \nu}$ \\
		\bottomrule
		\bottomrule
	 \end{tabular}}
	\label{table-operator}
\end{table}

One can project the correlation function decomposition to different helicity components of the parton by contracting the corresponding projection operators $\Gamma_{i j}$ listed in Table.~\ref{table-operator}, where we introduce a new superscript symbol $c(\bar{c})$ meaning $X^{c(\bar{c})} = X^{1} \pm i X^{2}$ or $X^{c(\bar{c})} = X^{x} \pm i X^{y}$.

Moreover, the correlator $\Phi^q \gamma^+$ about the quark helicity indices $i,j$ in Eq.~\eqref{correlator} describes the forward scattering amplitude of quark,
\begin{equation}\label{scattering-amplitude-quark}
		M_{ij}^q
		= {\rm Tr}(\Phi \gamma^+ \Gamma^q_{i j})
        = \int \left\langle P, S \left| \psi^{\dagger}_{+i}(0) \psi_{+j}(z^-) \right|P, S \right\rangle,
\end{equation}
where $S$ is used to stand for all the polarization information.
For further convenience, we also introduce the ``good" component of the gluon strength tensor,
\begin{eqnarray}
    G^{\,\,\, + \mu}_{[-]T} = \frac{1}{2} \left(G^{+\mu}_{\,\,\, T} -i \tilde{G}^{+\mu}_{\,\,\, T}\right), \, G^{\,\,\, +\mu}_{[+]T} = \frac{1}{2} \left(G^{+\mu}_{\,\,\, T} +i \tilde{G}^{+\mu}_{\,\,\, T}\right),
\end{eqnarray}
where $\tilde{G}^{+\mu}_{\,\,\, T} = \epsilon_{T}^{\mu \rho} G^{+ T}_{\;\;\; \rho}$ and $[\pm]$ stands for the gluon helicity.
Then, the gluon correlation function will be described by the distinct helicity gluon as
\begin{equation}\label{scattering-amplitude-gluon}
	\resizebox{.8\hsize}{!}{$\begin{split}
		M_{ij}^g
		= & \Phi^{g, \mu \nu} \Gamma_{i j,\mu \nu}^{g} \\
        = & \frac{1}{2 P^+} \int \left\langle P, S \left| G^{\,\, + c(\bar{c}) *}_{[i]}(0) G^{\,\, + c(\bar{c})}_{[j]} (z^-) \right|P, S \right\rangle,
	\end{split}$}
\end{equation}
where choosing $c$ or $\bar{c}$ depends on the corresponding helicity $[i(j)]$.
In the helicity space, the matrix $M_{ij}$ will be expressed as
\begin{widetext}
\begin{subequations}
\begin{equation}
	M^q_{i j}(x,S) = \frac{1}{2}
	\begin{pmatrix}\vspace{0.5em}
		f^q_1 + g^q_1S_L + f^q_{1LL} S_{LL} + g^q_{1LLL} S_{LLL} & h^q_1 S_{T}^c + h^q_{1LLT} S_{LLT}^c\\
		h^q_1 S_{T}^{\bar{c}} + h^q_{1LLT} S_{LLT}^{\bar{c}} & f^q_1 - g^q_1S_L + f^q_{1LL} S_{LL} - g^q_{1LLL} S_{LLL}
	\end{pmatrix},
\end{equation}
for quark, and
\begin{equation}
	M^g_{i j}(x,S) = \frac{1}{2}
	\begin{pmatrix}\vspace{0.5em}
		f^g_1 + g^g_1S_L + f^g_{1LL} S_{LL} + g^g_{1LLL} S_{LLL} & h^g_{1TT} S_{TT}^{x c}\\
		h^g_{1TT} S_{TT}^{x \bar{c}} & f^g_1 - g^g_1S_L + f^g_{1LL} S_{LL} - g^g_{1LLL} S_{LLL}
	\end{pmatrix},
\end{equation}
\end{subequations}
\end{widetext} 
for gluon. Furthermore, in the tensor product space of parton and hadron spins, the scattering amplitude has the forms,
\begin{widetext}
\begin{subequations}\label{spin-space}
	\begin{equation}\label{spin-space-quark}
		\begin{split}
			M^q_{\lambda' i, \lambda j} = & \int \left\langle P, \lambda' \left| \psi_{+i}^{\dagger}(0) \psi_{+j}(z^-) \right|P, \lambda \right\rangle
			=  \sum_n \left\langle P,\lambda' \left| \psi_{+i}^{\dagger} \right| P_n \right\rangle \left\langle P_n \left| \psi_{+j} \right| P, \lambda \right\rangle  \delta(P_n^+ - (1-x) P^+),
		\end{split}
	\end{equation}
    \begin{align}\label{spin-space-gluon}
        M^g_{\lambda' i, \lambda j} = & \frac{1}{2 P^+} \int
        \left\langle P, \lambda' \left| G^{\,\, + c (\bar{c}) *}_{[i]}(0) G^{\,\, + c (\bar{c})}_{[j]} (z^-) \right|P, \lambda \right\rangle =  \frac{1}{2 P^+} \sum_n \left\langle P,\lambda' \left| G^{\,\, + c (\bar{c}) *}_{[i]} \right| P_n \right\rangle \left\langle P_n \left| G^{\,\, + c (\bar{c})}_{[j]} \right| P, \lambda \right\rangle \notag \\
        & \times \delta(P_n^+ - (1-x) P^+).
    \end{align}
\end{subequations}
\end{widetext}

One can relate the scattering amplitude $M_{ij}$ in Eqs.~\eqref{scattering-amplitude-quark} and~\eqref{scattering-amplitude-gluon} and the spin space amplitude $M_{\lambda' i, \lambda j}$ in Eq.~\eqref{spin-space} by the hadron spin density matrix $\rho$,
\begin{equation}\label{matrix-relation}
  M_{ij} = M_{\lambda' i, \lambda j} \rho_{\lambda \lambda'}.
\end{equation}
The expression of the spin density matrix $\rho$ for the spin-3/2 particle~\cite{Zhao:2022lbw} is
\begin{equation}
    \rho = \frac{1}{4} \left( \bm{1} + \frac{4}{5} S^i \Sigma^i + \frac{2}{3} T^{ij} \Sigma^{ij} + \frac{8}{9} R^{i j k} \Sigma^{i j k} \right), 
\end{equation}
where $\Sigma$ is the four-dimensional polarization basis.
Solving Eq.~\eqref{matrix-relation},  with the space $\left.\left|\frac{3}{2} + \right.\right\rangle$, $\left.\left| \frac{1}{2} +\right.\right\rangle$, $\left.\left| -\frac{1}{2} +\right.\right\rangle$, $\left.\left| -\frac{3}{2} +\right.\right\rangle$, $\left.\left|\frac{3}{2} -\right.\right\rangle$, $\left.\left| \frac{1}{2} -\right.\right\rangle$, $\left.\left| -\frac{1}{2} -\right.\right\rangle$, and $\left.\left| -\frac{3}{2} -\right.\right\rangle$, one gets the spin space amplitude $M_{\lambda' i, \lambda j}$,
\begin{equation}\label{scattering-amplitude-spin-space}
	M_{\lambda' i,\lambda j}= \frac{1}{2}
  \begin{pmatrix} \vspace{0.5em}
	M_{\lambda' +,\lambda +} & M_{\lambda' +,\lambda -}\\
	M_{\lambda' -,\lambda +} & M_{\lambda' -,\lambda -}
  \end{pmatrix},
\end{equation}
where $M_{\lambda' i, \lambda j}$ is different between quark and gluon. Only a small quantity of terms are nonzero with the helicity conservation $\lambda' -i = \lambda -j $, so we will only write the specific form of all the nonzero matrix elements in the following.
For quark and gluon, the matrix elements with the same helicities between the absorbed and emitted partons, and they correspond to the unpolarized and longitudinal-parton PDFs, have the same forms,
\begin{equation}
\renewcommand{\arraystretch}{1.8}
    \resizebox{.85\hsize}{!}{$\begin{array}{c|c}
(\lambda', \lambda) & M^{q/g}_{\lambda' +, \lambda +}/M^q_{\lambda' -, \lambda -} \\ \hline 
(\frac{3}{2},\frac{3}{2})/(-\frac{3}{2},-\frac{3}{2}) &  f^{q/g}_1 + f^{q/g}_{1LL} + \frac{3}{2} g^{q/g}_1 + \frac{3}{10} g^{q/g}_{1LLL} \\
(\frac{1}{2},\frac{1}{2})/(-\frac{1}{2},-\frac{1}{2}) &  f^{q/g}_1 - f^{q/g}_{1LL} + \frac{1}{2} g^{q/g}_1 - \frac{9}{10} g^{q/g}_{1LLL} \\
(-\frac{1}{2},-\frac{1}{2})/(\frac{1}{2},\frac{1}{2}) &  f^{q/g}_1 - f^{q/g}_{1LL} - \frac{1}{2} g^{q/g}_1 + \frac{9}{10} g^{q/g}_{1LLL} \\
(-\frac{3}{2},-\frac{3}{2})/(\frac{3}{2},\frac{3}{2}) &  f^{q/g}_1 + f^{q/g}_{1LL} - \frac{3}{2} g^{q/g}_1 - \frac{3}{10} g^{q/g}_{1LLL}
\end{array}$},
\end{equation}
For the quark transversity-parton PDFs, the helicity conservation constrains $\lambda' - \lambda = i - j = \pm 1$,
\begin{equation}
\renewcommand{\arraystretch}{1.8}
    \begin{array}{c|c}
(\lambda', \lambda) & M^q_{\lambda' +, \lambda -}/M^q_{\lambda' -, \lambda +} \\ \hline 
(\frac{3}{2},\frac{1}{2})/(-\frac{3}{2},-\frac{1}{2}) &  \sqrt{3}h^q_1 + \frac{2 \sqrt{3}}{5} h^q_{1LLT} \\
(\frac{1}{2},-\frac{1}{2})/(-\frac{1}{2},\frac{1}{2}) &  2 h^q_1 - \frac{6}{5}h^q_{1LLT} \\
(-\frac{1}{2},-\frac{3}{2})/(\frac{1}{2},\frac{3}{2}) &  \sqrt{3}h^q_1 + \frac{2 \sqrt{3}}{5} h^q_{1LLT}
\end{array},
\end{equation}

For the gluon transversity-parton PDFs, the helicity conservation constrains $\lambda' -\lambda = i - j = \pm 2$, 
\begin{equation}
\renewcommand{\arraystretch}{1.8}
    \begin{array}{c|c}
(\lambda', \lambda) & M^g_{\lambda' +, \lambda -}/M^g_{\lambda' -, \lambda +} \\ \hline 
(\frac{3}{2},-\frac{1}{2})/(-\frac{3}{2},\frac{1}{2}) &  2\sqrt{3}h^g_{1TT} \\
(\frac{1}{2},-\frac{3}{2})/(-\frac{1}{2},\frac{3}{2}) &  2\sqrt{3}h^g_{1TT}
\end{array},
\end{equation}
In these above equations, the index $T$ of the PDFs implies the spin flip. Therefore, one $T$ appears in the quark PDFs and two $T$s in the gluon PDFs due to different spins.

\paragraph*{Relations between helicity amplitudes and PDFs.}
Observing the matrix $M_{\lambda' i, \lambda j}$, one finds the definition is the same with the corresponding helicity amplitudes in the forward limit,
\begin{subequations}\label{helicity-amplitudes}
	\begin{equation}
        \mathcal{A}^q_{\lambda' i, \lambda j} = M^q_{\lambda' i, \lambda j} = \int 
		\left\langle P, \lambda' \left| \bar{\psi}(0) \gamma^+ \Gamma^q_{i j} \psi(z^-) \right|P, \lambda \right\rangle.
	\end{equation}
for quark, and
 	\begin{equation}
        \mathcal{A}^g_{\lambda' i, \lambda j} = M^g_{\lambda' i, \lambda j} = \frac{1}{P^+}\int
		\left\langle P, \lambda' \left| G^{+ \mu}(0) \Gamma^g_{i j, \mu \nu} G^{\nu +}(z^-) \right|P, \lambda \right\rangle.
	\end{equation}
\end{subequations}
for gluon.
Analyzing the helicity amplitude $\mathcal{A}_{\lambda' i, \lambda j}$ allows for a deeper understanding of its properties and enables the derivation of its relations with the distribution functions.
From the hermiticity, parity and time reversal transformations, the helicity amplitudes~\eqref{helicity-amplitudes} in the forward limit satisfy the constraints~\cite{Diehl:2003ny,Fu:2022bpf,Fu:2024kfx},
  \begin{equation}
	\mathcal{A}_{\lambda' i, \lambda, j} = \mathcal{A}_{\lambda j, \lambda' i}, \quad
	\mathcal{A}_{-\lambda' -i, -\lambda, -j} = \mathcal{A}_{\lambda j, \lambda' i}.  
\end{equation}
Meanwhile, we have concluded that the helicity amplitudes are real and only the amplitudes with helicity conservation $\lambda'-\lambda = i - j$ are nonzero in the forward limit~\cite{Fu:2022bpf,Fu:2024kfx}.
Therefore, there are only six independent amplitudes and PDFs for quark in the spin-3/2 particle and five for gluon.
The helicity amplitude matrix $\mathcal{A}_{\lambda' i, \lambda j}$ in the tensor space of the parton and hadron spins like $M_{\lambda' i, \lambda j}$ are shown as
\begin{widetext}
\begin{subequations}\label{amplitude-matrix}
	\begin{equation}
		\mathcal{A}^q=\begin{pmatrix} 
			\mathcal{A}^q_{\frac{3}{2} +, \frac{3}{2} +} & 0 & 0 & 0 & 0 & \mathcal{A}^q_{\frac{1}{2} -, \frac{3}{2} +} & 0 & 0\\
			0 & \mathcal{A}^q_{\frac{1}{2} +, \frac{1}{2} +} & 0 & 0 & 0 & 0 & \mathcal{A}^q_{-\frac{1}{2} -, \frac{1}{2} +} & 0\\
			0 & 0 & \mathcal{A}^q_{-\frac{1}{2} +, -\frac{1}{2} +} & 0 & 0 & 0 & 0 & \mathcal{A}^q_{\frac{1}{2} -, \frac{3}{2} +}\\
			0 & 0 & 0 & \mathcal{A}^q_{-\frac{3}{2} +, -\frac{3}{2} +} & 0 & 0 & 0 & 0\\
			0 & 0 & 0 & 0 & \mathcal{A}^q_{-\frac{3}{2} +, -\frac{3}{2} +} & 0 & 0 & 0\\
			\mathcal{A}^q_{\frac{1}{2} -, \frac{3}{2} +} & 0 & 0 & 0 & 0 & \mathcal{A}_{-\frac{1}{2} +, -\frac{1}{2} +} & 0 & 0\\
			0 & \mathcal{A}^q_{-\frac{1}{2} -, \frac{1}{2} +} & 0 & 0 & 0 & 0 & \mathcal{A}^q_{\frac{1}{2} +, \frac{1}{2} +} & 0 \\
			0 & 0 & \mathcal{A}^q_{\frac{1}{2} -, \frac{3}{2} +} & 0 & 0 & 0 & 0 & \mathcal{A}^q_{\frac{3}{2} +, \frac{3}{2} +} \\
		\end{pmatrix},
	\end{equation}	
    \begin{equation}
		\mathcal{A}^g=\begin{pmatrix} 
			\mathcal{A}^g_{\frac{3}{2} +, \frac{3}{2} +} & 0 & 0 & 0 & 0 & 0 & \mathcal{A}^g_{-\frac{1}{2} -, \frac{3}{2} +} & 0\\
			0 & \mathcal{A}^g_{\frac{1}{2} +, \frac{1}{2} +} & 0 & 0 & 0 & 0 & 0 & \mathcal{A}^g_{-\frac{1}{2} -, \frac{3}{2} +}\\
			0 & 0 & \mathcal{A}^g_{-\frac{1}{2} +, -\frac{1}{2} +} & 0 & 0 & 0 & 0 & 0\\
			0 & 0 & 0 & \mathcal{A}^g_{-\frac{3}{2} +, -\frac{3}{2} +} & 0 & 0 & 0 & 0\\
			0 & 0 & 0 & 0 & \mathcal{A}^g_{-\frac{3}{2} +, -\frac{3}{2} +} & 0 & 0 & 0\\
			0 & 0 & 0 & 0 & 0 & \mathcal{A}^g_{-\frac{1}{2} +, -\frac{1}{2} +} & 0 & 0\\
			\mathcal{A}^g_{-\frac{1}{2} -, \frac{3}{2} +} & 0 & 0 & 0 & 0 & 0 & \mathcal{A}^g_{\frac{1}{2} +, \frac{1}{2} +} & 0 \\
			0 & \mathcal{A}^g_{-\frac{1}{2} -, \frac{3}{2} +} & 0 & 0 & 0 & 0 & 0 & \mathcal{A}^g_{\frac{3}{2} +, \frac{3}{2} +}
		\end{pmatrix},
	\end{equation}
\end{subequations}
\end{widetext}
for quark and gluon, respectively,
where we just write the nonzero and independent terms.

Comparing Eq.~\eqref{scattering-amplitude-spin-space} and Eq.~\eqref{amplitude-matrix}, one gives the equalities between the helicity amplitudes and PDFs by the one-to-one correspondence of matrix elements.
Solving the equations, one obtains the distribution function definitions from the helicity amplitudes. The unpolarized and longitudinal-parton PDFs have the common definitions,
\begin{align}
    f_1^{q/g} = & \left(\mathcal{A}^{q/g}_{\frac{3}{2} +, \frac{3}{2} +} + \mathcal{A}^{q/g}_{-\frac{3}{2} +, -\frac{3}{2} +} \right. \notag \\ & \,\,\left. + \mathcal{A}^{q/g}_{\frac{1}{2} +, \frac{1}{2} +} + \mathcal{A}^{q/g}_{-\frac{1}{2} +, -\frac{1}{2} +}\right)\Big/2, \notag \\
    f_{1LL}^{q/g} = & \left[\left( \mathcal{A}^{q/g}_{\frac{3}{2} +, \frac{3}{2} +} + \mathcal{A}^{q/g}_{-\frac{3}{2} +, -\frac{3}{2} +} \right) \right. \notag \\ & \,\, \left. -\left(   \mathcal{A}^{q/g}_{\frac{1}{2} +, \frac{1}{2} +} + \mathcal{A}^{q/g}_{-\frac{1}{2} +, -\frac{1}{2} +} \right) \right]\Big/ 2,\\
    g_{1}^{q/g} = & \left[3\left( \mathcal{A}^{q/g}_{\frac{3}{2} +, \frac{3}{2} +} - \mathcal{A}^{q/g}_{-\frac{3}{2} +, -\frac{3}{2} +} \right) \right. \notag \\  & \,\, \left. + \left(   \mathcal{A}^{q/g}_{\frac{1}{2} +, \frac{1}{2} +} - \mathcal{A}^{q/g}_{-\frac{1}{2} +, -\frac{1}{2} +} \right)\right]\Big/5, \notag \\
    g_{1LLL}^{q/g} &= \left[\left( \mathcal{A}^{q/g}_{\frac{3}{2} +, \frac{3}{2} +} - \mathcal{A}^{q/g}_{-\frac{3}{2} +, -\frac{3}{2} +} \right) \right. \notag \\ & \,\, \left.  - 3\left(\mathcal{A}^{q/g}_{\frac{1}{2} +, \frac{1}{2} +} - \mathcal{A}^{q/g}_{-\frac{1}{2} +, -\frac{1}{2} +} \right)\right]\Big/3.\notag 
\end{align}
Because the transversity-parton PDFs relate to the spin flip, the quark and gluon PDFs have the different forms. The quark transversity-parton PDFs are defined as
\begin{equation}
\begin{split}
    h_1^{q} & = \frac{2}{5}\left(\sqrt{3} \mathcal{A}^{q}_{\frac{1}{2} -, \frac{3}{2} +} + \mathcal{A}^{q}_{-\frac{1}{2} -, \frac{1}{2} +}\right), \\
	h_{1LLT}^{q} & = \frac{1}{3}\left(2\sqrt{3} \mathcal{A}^{q}_{\frac{1}{2} -, \frac{3}{2} +} - 3 \mathcal{A}^{q}_{-\frac{1}{2} -, \frac{1}{2} +}\right),
\end{split}
\end{equation}
and the gluon transversity-parton PDF is defined as
\begin{equation}
    h^g_{1TT} = \frac{1}{ \sqrt{3}}\mathcal{A}^g_{-\frac{1}{2} -, \frac{3}{2} +}.
\end{equation}
With the increase of spin, there are more independent PDFs.
The PDFs $f_{1LL}^{q/g}$ and $h_{1TT}^g$ only appear when the spin $s \geq 1$ and $g_{1LLL}^{q/g}$ and $h_{1LLT}^q$ only appear in the spin-3/2 and higher spin particle.
\paragraph*{Positivity constraints.}

Furthermore, being the square of a scattering amplitude as explained in Eq.~\eqref{spin-space}, the matrix $M_{\lambda' i,\lambda j}$ in Eq.~\eqref{spin-space} and helicity amplitudes in Eq.~\eqref{amplitude-matrix} need to satisfy the positive definiteness, which constrains the helicity amplitudes,
\begin{subequations}\label{eq:amplitude-relation}
    \begin{equation}
	    \mathcal{A}^{q/g}_{a + , a + } \ge 0 \quad \text{with} \quad a = \frac{3}{2}, \frac{1}{2}, -\frac{1}{2}, -\frac{3}{2},
\end{equation}
for quark and gluon,
\begin{equation}
		\mathcal{A}^{q}_{\frac{3}{2} +, \frac{3}{2} +} \mathcal{A}^{q}_{-\frac{1}{2} +, -\frac{1}{2} +} \ge  \left|\mathcal{A}^{q}_{\frac{1}{2} -, \frac{3}{2} +}\right|^2, \,
		\mathcal{A}^{q}_{\frac{1}{2} +, \frac{1}{2} +} \ge  \left|\mathcal{A}^{q}_{-\frac{1}{2} -, \frac{1}{2} +}\right|,
\end{equation}
for quark, and
\begin{equation}
    \mathcal{A}^g_{\frac{3}{2} +, \frac{3}{2} +} \mathcal{A}^g_{-\frac{1}{2} -, -\frac{1}{2} -} \ge  \left|\mathcal{A}^{g}_{-\frac{1}{2} -, \frac{3}{2} +}\right|^2,
\end{equation}
for gluon.
\end{subequations}

We can summarize the rule for any spin-$s$ particle. The helicity amplitudes in the forward limit have the constraints,
	\begin{align}
		\mathcal{A}_{a +, a +} \ge 0 \quad & \text{with} \quad a = s, s-1 \cdots -s, \notag \\
		\mathcal{A}_{a +, a +} \mathcal{A}_{b -, b -} \ge \mathcal{A}_{b -, a +}^2 \quad & \text{with} \quad a = s, s-1 \cdots > 0,\notag 
	\end{align} 
where $a-b = 1$ for quark and $a - b = 2$ for gluon from the helicity conservation.
For the spin-$s$ particle, one can obtain $2 s + 1 + \lceil s \rceil$ relations for quark and $2 s + 1 + \lfloor s \rfloor$ relations for gluon, where $\lceil s \rceil$ and $\lfloor s \rfloor$ are the round up and round down functions.

Meanwhile, from the positive definiteness of the matrix $M_{\lambda' i,\lambda j}$ or the helicity amplitude inequalities in Eq.~\eqref{eq:amplitude-relation}, one obtains the bounds among the PDFs for the spin-3/2 particle,
\begin{subequations}\label{soffer-bound_2}
    \begin{eqnarray}
       && f_1^{q/g} + f_{1LL}^{q/g} \ge  \left| \frac{3}{2} g_1^{q/g} + \frac{3}{10} g_{1LLL}^{q/g} \right|, \\
    &&    f_1^{q/g} - f_{1LL}^{q/g} \ge  \left|  \frac{1}{2} g_1^{q/g} - \frac{9}{10} g_{1LLL}^{q/g} \right|,
    \end{eqnarray}
for quark and gluon,
\begin{widetext}
\begin{eqnarray}
    \left(f_1^{q} + f_{1LL}^{q} + \frac{3}{2} g_1^{q} + \frac{3}{10} g_{1LLL}^{q}\right) \left(f_1^{q} - f_{1LL}^{q} - \frac{1}{2} g_1^{q} + \frac{9}{10} g_{1LLL}^{q}\right) & \ge & 3 \left( h_1^{q} + \frac{2}{5} h_{1LLT}^{q} \right)^2 , \\
    f_1^{q} - f_{1LL}^{q} + \frac{1}{2} g_1^{q} - \frac{9}{10} g_{1LLL}^{q} & \ge &  \left| 2 h_1^{q} - \frac{6}{5} h_{1LLT}^{q} \right|,
\end{eqnarray}
for quark, and
\begin{eqnarray}
	\left(f_1^g + f_{1LL}^g + \frac{3}{2} g_1^g + \frac{3}{10} g_{1LLL}^g\right) \left(f_1^g - f_{1LL}^g + \frac{1}{2} g_1^g - \frac{9}{10} g_{1LLL}^g\right) & \ge & 12 \left(h_{1TT}^{g}\right)^2 ,
\end{eqnarray}
\end{widetext}
\end{subequations}
for gluon.
The resulting bounds generalize the classical Soffer bound to the spin-3/2 case and provide essential theoretical constraints for the model calculation or global QCD analysis of the high-spin hadron structure. They will serve as a necessary consistency check for future extractions of spin-3/2 PDFs from lattice QCD or experimental data.

In the non-forward limit, the initial and final states have the different momenta and the quark GPD is defined as
\begin{widetext}
	\begin{equation}
		\int \left\langle p', S \left| \psi^\dagger_{+i}(0) \psi_{+j}(z^-) \right| p, S \right\rangle
		= \sum_n \left\langle p', S \left| \psi^\dagger_{+i}(0) \right| P_n\right\rangle \left\langle P_n \left| \psi_{+j}(0) \right| p, S \right\rangle \delta\left( P_n^+ - p^+ + x \bar{P}^+ \right),
	\end{equation}
	where $\bar{P}=\frac{p+p'}{2}$, and which will reduce to the PDF when the initial and final state momenta become the same. According to the Cauchy-Schwartz inequality~\cite{Pire:1998nw},
	\begin{equation}
		\sum_n \left| \left\langle p, S \left| \psi^\dagger_{+i}(0) \right| P_n \right\rangle \pm a \left\langle p',S \left| \psi_{+j}(0) \right| P_n \right\rangle \right|^2 \delta\left( P_n^+ - (1-x)p^+ \right) \ge 0,
	\end{equation}
	where $a$ is a arbitrary coefficient, one obtains
	\begin{equation}\label{CS-2}
		\begin{split}
			& \sum_n  \left\langle p, S \left| \psi^\dagger_{+i}(0) \right| P_n \right\rangle \left\langle P_n \left| \psi_{+i}(0) \right| p, S \right\rangle \delta\left( P_n^+ - (1-x)p^+ \right)
			+  \sum_n \left| a \right|^2 \left\langle p', S \left| \psi^\dagger_{+j}(0) \right| P_n \right\rangle\left\langle P_n \left| \psi_{+j}(0) \right| p', S \right\rangle \delta\left( P_n^+ - (1-x)p^+ \right) \\ 
			& \ge \, 2 Re \sum_n a \left\langle p', S \left| \psi^\dagger_{+i}(0) \right| P_n\right\rangle \left\langle P_n \left| \psi_{+j}(0) \right| p, S \right\rangle \delta\left( P_n^+ - (1-x)p^+ \right).
		\end{split}
	\end{equation}
\end{widetext}
In the corresponding definitions of the PDFs and GPDs, the parton momentum fraction has the different definition, $x_1 = x = \frac{k^+}{p^+}$, $x_2 = \frac{k^+}{p'^+}$ and $x_3 = \frac{k^+}{\bar{P}^+}$, which have the relation,
\begin{equation}
	p^+ - x_1 p^+ = p'^+ - x_2 p'^+ = p^+ - x_3 \bar{P}^+.
\end{equation}
Then, Eq.~\eqref{CS-2} becomes the relation between the PDF and GPD of the unpolarized quark,
\begin{equation}
	f_1^q(x_1) + \left| a \right|^2  f_1^q(x_2) \ge 2 Re \left[ a f_1^q(x_3, \xi, t)\right].
\end{equation}
According to the analogical process with the quark, one obtains the gluon relation,
\begin{equation}
	\left( 1+\xi \right)f_1^g(x_1) + \left( 1-\xi \right)\left| a \right|^2  f_1^g(x_2) \ge 2 Re \left[ a f_1^g(x_3, \xi, t)\right].
\end{equation}
These relations imply that the spin-averaged quark GPD and gluon GPD are constrained by the corresponding PDFs.
If we set the initial and final particles with different spins, one obtains the bounds on the transition GPD from the initial and final particle PDFs. For example, the $N-\Delta$ transition GPD can be estimated and constrained by the nucleon and $\Delta$ PDFs~\cite{Pobylitsa:2001nt,Kirch:2005in,Kim:2024hhd}.
These positivity bounds provide extra restrictions for model inputs and reference for the PDF and GPD extraction from the experiments~\cite{CLAS:2023akb,Diehl:2024bmd,Fernando:2025xzv}.
Experimentally, $^7$Li is a stable and promising spin-$\frac{3}{2}$ target to measure the PDFs, transverse-momentum-dependent parton distributions (TMDs), and GPDs by deep inelastic scattering (DIS), semi-inclusive deep inelastic scattering (SIDIS), deeply-virtual Compton scattering (DVCS), and Drell-Yan processes~\cite{Anderle:2021wcy}. 
For example, it is possible to measure the PDFs by the Drell-Yan process with the $^7$Li target in the Fermilab SpinQuest experiment~\cite{Keller}.

\paragraph*{Summary.}

In this work, we established the foundational positivity constraints for the leading-twist quark and gluon PDFs of a spin-3/2 hadron, thereby generalizing the well-known Soffer bound from spin-1/2 and spin-1 to the higher-spin regime. The derivation is founded in a clear physical principle: the antiparton-hadron forward scattering amplitude $M_{\lambda' i, \lambda j}$ in the tensor space of the parton and hadron spins, and it is expressed in terms of the PDFs and must form a positive definite matrix.

To achieve this, we first defined the complete set of six independent quark PDFs $ (f^{q}_1,\, g^{q}_1,\, h^{q}_1,\, f^{q}_{1LL},\, g^{q}_{1LLL},\, h^{q}_{1LLT}) $ and five independent gluon PDFs $ (f^{g}_1,\, g^{g}_1,\, f^{g}_{1LL},\, h^{g}_{1TT},\, g^{g}_{1LLL}) $ for a spin-3/2 particle within the light-cone correlation function framework. A crucial step is the connection established between these PDFs and the antiparton-hadron forward scattering amplitude $M_{\lambda' i, \lambda j}$.
Meanwhile, we obtained the definitions of the PDFs from helicity amplitude $\mathcal{A}_{\lambda' i, \lambda j}$, which has the same definition with $M_{\lambda' i, \lambda j}$.
Positive definiteness of the antiparton-hadron forward scattering amplitude constrains the helicity amplitudes by \(2s + 1+\lceil s \rceil\) inequalities for quark and \(2s + 1+\lfloor s \rfloor\) inequalities for gluon. Meanwhile, positive definiteness determines the bounds on the quark and gluon PDFs of the spin-3/2 particle.
Moreover, we also obtained the quark and gluon constraints between the PDFs and GPDs.
The derived bounds are essential for the future of high-spin hadron structure studies.
These inequalities provide the rigorous mathematical boundaries that any physically viable model or extraction of these distributions must obey.

\paragraph*{Acknowledgments.}

This work is partly supported by the National Key R\&D Program of China under Grant No. 2023YFA1606703, and by the National Natural Science Foundation of China under Grants Nos. 12375142, 12575094, 12435007, 12361141819 and 12447121. It is also supported by the Gansu Province Postdoctoral Foundation.

\bibliographystyle{unsrt}
\bibliography{refs1}
\end{document}